\documentclass[a4paper,11pt]{article}

\usepackage{booktabs}

\usepackage{amsthm}
\usepackage{amsmath}
\usepackage{amssymb}
\usepackage{graphicx}
\usepackage{microtype} 
\usepackage[round]{natbib}
\usepackage{bbm}
\usepackage{textcomp}

\usepackage{color}
\usepackage{framed}
\usepackage{comment}
\definecolor{shadecolor}{gray}{0.9}
\usepackage{dsfont}

\specialcomment{extra}{\begin{shaded}}{\end{shaded}}

\theoremstyle{plain}  %default 

\theoremstyle{definition}

\theoremstyle{remark}

\newcommand{\diff}{\,\mathrm{d}}
\newcommand{\E}{\mathbb{E}}

\newcommand{\ES}{\operatorname{ES}}

\newcommand{\R}{\mathbb{R}}

\newcommand{\argmin}{\operatorname{arg\,min}}
\newcommand{\VaR}{\operatorname{VaR}}
\newcommand{\one}{\mathds{1}}

\newcommand{\cN}{\mathcal{N}}

\begin{document}

\title{Expected Shortfall is jointly elicitable with Value at Risk --
Implications for backtesting}

\author{Tobias Fissler\thanks{University of Bern, Department of
  Mathematics and Statistics, Institute of Mathematical Statistics and
  Actuarial Science, Sidlerstrasse 5, 3012 Bern, Switzerland, e-mail:
  \texttt{tobias.fissler@stat.unibe.ch};
  \texttt{johanna.ziegel@stat.unibe.ch}} \and Johanna
  F.~Ziegel\footnotemark[1] \and Tilmann Gneiting\thanks{Heidelberg
  Institute for Theoretical Studies and Karlsruhe Institute of Technology, 
  HITS gGmbH, Schlo{\ss}-Wolfsbrunnenweg 35, 69118 Heidelberg, Germany, e-mail: 
  \texttt{tilmann.gneiting@h-its.org}}}

\maketitle

\begin{abstract}
In this note, we comment on the relevance of elicitability for backtesting risk measure estimates. In particular, we propose 
the use of Diebold-Mariano tests, and show how they can be implemented for Expected Shortfall (ES), based on the recent 
result of \citet{FisslerZiegel2015} that ES is jointly elicitable with Value at Risk.
\end{abstract}

There continues to be lively debate about the appropriate choice of a
quantitative risk measure for regulatory purposes or internal risk
management.  In this context, it has been shown by \citet{Weber2006}
and \citet{Gneiting2011} that Expected Shortfall (ES) is not
elicitable.  Specifically, there is no strictly consistent scoring (or
loss) function $S : \R^2 \to \R$ such that, for any random variable
$X$ with finite mean, we have
\[
\ES_{\alpha}(X) = \argmin_{e \in \R} \E[S(e,X)].
\]
Recall that ES of $X$ at level $\alpha \in (0,1)$ is defined as
\[
\ES_{\alpha}(X) = \frac{1}{\alpha} \int_0^\alpha \text{VaR}_\beta(X) \diff \beta,
\]
where Value at Risk (VaR) is given by $\VaR_{\alpha}(X) = \inf\{ x \in
\R \colon \mathbb{P}(X \le x) \ge \alpha\}$.  In contrast, VaR at
level $\alpha \in (0,1)$ is elicitable for random variables with a
unique $\alpha$-quantile.  The possible strictly consistent scoring
functions for VaR are of the form 
\begin{equation} \label{eq:S_V}
S_V(v,x) = (\one\{x \le v\} - \alpha)(G(v) - G(x)), 
\end{equation}
where $G$ is a strictly increasing function.

However, it turns out that ES is elicitable of higher order in the
sense that the pair (VaR$_\alpha$, ES$_\alpha$) is jointly elicitable.
Indeed, we have that
\[
(\VaR_{\alpha}(X),\ES_{\alpha}(X)) = \argmin_{(v,e) \in \R^2} \E[S_{V,E}(v,e,X)],
\]
where possible choices of $S_{V,E}$ are given by
\begin{align} \label{eq:S_VE}
S_{V,E}(v,e,x) 
& = (\one\{x\le v\} -\alpha )(G_1(v) - G_1(x)) \\
& \quad + \frac{1}{\alpha} G_2(e)\one\{x \le v\}(v - x) 
  + G_2(e)(e - v) - \mathcal{G}_2(e),
\nonumber
\end{align}
with $G_1$ and $G_2$ being strictly increasing continuously
differentiable functions such that the expectation $\E[G_1(X)]$
exists, $\lim_{x \to -\infty} G_2(x) = 0$ and $\mathcal G_2' = G_2$;
see \citet[Corollary 5.5]{FisslerZiegel2015}.  One can nicely see the
structure of $S_{V,E}$: The first summand in \eqref{eq:S_VE} is
exactly a strictly consistent scoring function for VaR$_\alpha$ given
at \eqref{eq:S_V} and hence only depends on $v$, whereas the second
summand cannot be split into a part depending only on $v$ and one
depending only on $e$, respectively, hence illustrating the fact that
ES$_\alpha$ itself is not elicitable.  A possible choice for $G_1$ and
$G_2$ is $G_1(v) = v$ and $G_2(e) = \exp(e)$.
\cite{AcerbiSzekely2014} proposed a scoring function for the pair
(VaR$_\alpha$, ES$_\alpha$) under the additional assumption that there
exists a real number $w$ such that $\text{\rm ES}_{\alpha}(X) > w \,
\text{\rm VaR}_{\alpha}(X)$ for all assets $X$ under
consideration. Despite encouraging simulation results, there is
currently no formal proof available of the strict consistency of their
proposal.  In contrast, the scoring functions given at \eqref{eq:S_VE}
do not require additional assumptions, and it has been formally proven
that they provide a class of strictly consistent scoring functions.

The lack of elicitability of ES (of first order) has led to a lively
discussion about whether or not and how it is possible to backtest ES
forecasts; see, for example, \citet{AcerbiSzekely2014},
\citet{Carver2014}, and \citet{EmmerETAL2013}.  It is generally
accepted that elicitability is useful for model selection, estimation,
generalized regression, forecast comparison, and forecast ranking.
Having provided strictly consistent scoring functions for
(VaR$_\alpha$, ES$_\alpha$), we take the opportunity to comment on the
role of elicitability in backtesting.

The traditional approach to backtesting aims at model verification.
To this end, one tests the null hypothesis: 
\[ 
H_0^C : \text{``The risk measure estimates at hand are correct.''}  
\] 
Specifically, suppose we have sequences $(x_t)_{t=1,\ldots,N}$ and
$(v_t,e_t)_{t=1,\ldots,N}$, where $x_t$ is the realized value of the
asset at time point $t$, and $v_t$ and $e_t$ denote the estimated
VaR$_\alpha$ and ES$_\alpha$ given at time $t-1$ for time point $t$,
respectively.  A backtest uses some test statistic $T_1$, which is a
function of $(v_t,e_t,x_t)_{t=1,\ldots,N}$, such that we know the
distribution of $T_1$ (at least approximately) if the null hypothesis
of correct risk measure estimates holds.  If we reject $H_0^C$ at some
small level, the model or the estimation procedure for the risk
measure is deemed inadequate.  For this approach of model
verification, elicitability of the risk measure is not relevant, as
pointed out by \citet{AcerbiSzekely2014} and \citet{Davis2014}.
However, tests of this type can be problematic in regulatory practice,
notably in view of the anticipated revised standardised approach
\citep[pp.~5--6]{BIS2013}, which ``should provide a credible fall-back
in the event that a bank's internal market risk model is deemed
inadequate''.  If the internal model fails the backtest, the
standardised approach may fail the test, too, and in fact it might be
inferior to the internal model.  Generally, tests of the hypothesis
$H_0^C$ are not aimed at, and do not allow for, model comparison and
model ranking.

Alternatively, one could use the following null hypothesis in
backtesting: 
\[
H_0^- : \parbox{9.8cm}{\centering``The risk measure estimates at hand are \emph{at least as
good} as the ones from the standard procedure.''}
\]
Here, the standard procedure could be a method specified by the
regulator, or it could be a technique that has proven to yield good
results in the past.  Specifically, let us write
$(v_t^*,e_t^*)_{t=1,\ldots,N}$ for the sequence of VaR$_\alpha$ and
ES$_\alpha$ estimates by the standard procedure.  Making use of the
elicitability of (VaR$_\alpha$, ES$_\alpha$), we take one of the
scoring functions $S_{V,E}$ given at \eqref{eq:S_VE} to define the
test statistic
\begin{equation}  \label{eq:DM} 
T_2 = \frac{\bar{S}_{V,E}-\bar{S}_{V,E}^*}{\sigma_N},
\end{equation} 
where
\[
\bar{S}_{V,E} = \frac{1}{N} \sum_{t=1}^N S_{V,E}(v_t,e_t,x_t),  
\quad 
\bar{S}_{V,E}^* = \frac{1}{N} \sum_{t=1}^N S_{V,E}(v_t^*,e_t^*,x_t),
\]
and $\sigma_N$ is a suitable estimate of the respective standard
deviation.  Under $H_0^-$, the test statistic $T_2$ has expected value
less than or equal to zero.  Following the lead of
\citet{DieboldMariano1995}, comparative tests that are based on the
asymptotic normality of the test statistics $T_2$ have been employed
in a wealth of applications.

Under both $H_0^C$ and $H_0^-$, the backtest is passed if the null
hypothesis fails to be rejected.  However, as
\citet[p.~16]{Fisher1949} noted, ``the null hypothesis is never proved
or established, but it is possibly disproved, in the course of
experimentation.''  In other words, a passed backtest does not imply
the validity of the respective null hypothesis.  Passing the backtest
simply means that the hypothesis of correctness ($H_0^C$) or
superiority ($H_0^-$), respectively, could not be falsified.

\begin{figure}[t] 
\centering
\includegraphics[width=0.8\textwidth]{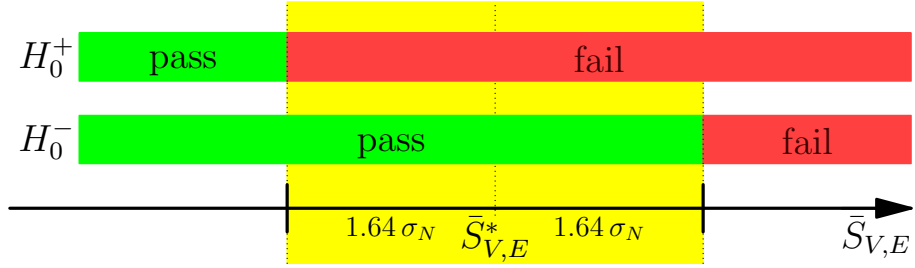} 
\caption{Decisions taken in comparative backtests under the null
  hypotheses $H_0^-$ and $H_0^+$ at level $0.05$.  In the
  yellow region the approaches entail distinct decisions.
  \label{fig:1}}
\end{figure}

In the case of comparative backtests, a more conservative approach
could be based on the following null hypothesis:
\[ 
H_0^+ : \parbox{9.8cm}{\centering``The risk measure estimates at hand are
\emph{at most as good} as the ones from the standard procedure."}  
\]
This can also be tested using the statistic $T_2$ in \eqref{eq:DM},
which has expected value greater than or equal to zero under $H_0^+$.
The backtest now is passed when $H_0^+$ is rejected.  The decisions
taken in comparative backtesting under $H_0^-$ and $H_0^+$ are
illustrated in Figure \ref{fig:1}, where the colors relate to the
three-zone approach of the \citet[pp.~103--108]{BIS2013}.  In
regulatory practice, the distinction between Diebold-Mariano tests
under the two hypotheses amounts to a reversed onus of proof.  In the
traditional setting, it is the regulator's burden to show that the
internal model is incorrect.  In contrast, if a backtest is passed
when $H_0^+$ is rejected, banks are obliged to demonstrate the
superiority of the internal model.  Such an approach to backtesting
may entice banks to improve their internal models, and is akin to
regulatory practice in the health sector, where market authorisation
for medicinal products hinges on comparative clinical trials.  In the
health context, decision-making under $H_0^-$ corresponds to
equivalence or non-inferiority trials, which are ``not conservative in
nature, so that many flaws in the design or conduct of the trial will
tend to bias the results'', whereas ``efficacy is most convincingly
established by demonstrating superiority'' under $H_0^+$
\citep[p.~17]{ICHE9}.  Technical detail is available in a specialized
strand of the biomedical literature; for a concise review, see
\citet{Lesaffre2008}.

We now give an illustration in the simulation setting of
\citet{GneitingETAL2007}.  Specifically, let $(\mu_t)_{t=1,\dots,N}$
be a sequence of independent standard normal random variables.
Conditional on $\mu_t$, the return $X_t$ is normally distributed with
mean $\mu_t$ and variance 1, denoted $\cN(\mu_t,1)$.  Under our
Scenario A, the standard method for estimating risk measures uses the
unconditional distribution $\cN(0,2)$ of $X_t$, whereas the internal
procedure takes advantage of the information contained in $\mu_t$ and 
uses the conditional distribution $\cN(\mu_t,1)$.  Therefore,
\[
\left( v_t, e_t \right)
= \left( \VaR_{\alpha}(\cN(\mu_t,1)), \ES_{\alpha}(\cN(\mu_t,1)) \right) 
= \left( \mu_t + \Phi^{-1}(\alpha), \,
         \mu_t - \frac{1}{\alpha} \, \varphi(\Phi^{-1}(\alpha)) \right) 
\]
and 
\[
\left( v_t^*, e_t^* \right) 
= \left( \VaR_{\alpha}(\cN(0,2)), \ES_{\alpha}(\cN(0,2)) \right) 
= \left( \sqrt{2} \, \Phi^{-1}(\alpha), \,  
         - \frac{\sqrt{2}}{\alpha} \, \varphi(\Phi^{-1}(\alpha)) \right) \! ,
\]
where $\varphi$ and $\Phi$ denote the density and the cumulative
distribution function of the standard normal distribution,
respectively.  Under Scenario B, the roles of the standard method and
the internal procedure are interchanged.

We use sample size $N = 250$ and repeat the experiment 10,000 times.
As tests of traditional type, we consider the coverage test for
$\VaR_{0.01}$ described by the \citet[pp.~103--108]{BIS2013} and the
generalized coverage test for $\ES_{0.025}$ proposed by
\citet{CostanzinCurran2015}.  As shown by
\citet{CliftCostanzinETAL2015}, the latter performs similarly to the
approaches of \citet{Wong2008} and \citet{AcerbiSzekely2014}, but is
easier to implement.  The outcome of the test is structured into
green, yellow, and red zones, as described in the aforementioned
references.  For the comparative backtest for $(\VaR_{0.025},
\ES_{0.025})$, we use the functions $G_1(v) = v$ and $G_2(e) =
\exp(e)/(1 + \exp(e))$ in \eqref{eq:S_VE} and define the zones as
implied by Figure \ref{fig:1}.  Finally, our comparative backtest for
$\VaR_{0.01}$ uses the function $G(v) = v$ in \eqref{eq:S_V}, which is
equivalent to putting $G_1(v) = v$ and $G_2(e) = 0$ in
\eqref{eq:S_VE}.  For $\sigma_N$ in the test statistic $T_2$ in
\eqref{eq:DM} we use the standard estimator.

\begin{table}[t] 

\caption{Percentage of decisions in the green, yellow, and red zone in
  traditional and comparative backtests.  Under Scenario A, a decision
  in the green zone is desirable and in the joint interest of banks
  and regulators.  Under Scenario B, the red zone corresponds to a
  decision in the joint interest of all stakeholders.  \label{tab:1}}
\begin{center}
\begin{tabular}{|l|l|ccc|}
\toprule
{\bf Scenario A} & & {\bf Green} & Yellow & Red \\
\midrule
Traditional & $\VaR_{0.01}$              & 89.35 & 10.65 & \hphantom{0}0.00 \\
Traditional & $\ES_{0.025}$              & 93.62 & \hphantom{0}6.36 & \hphantom{0}0.02 \\
Comparative & $\VaR_{0.01}$              & 88.23 & 11.77 & \hphantom{0}0.00 \\
Comparative & $(\VaR_{0.025},\ES_{0.025})$ & 87.22 & 12.78 & \hphantom{0}0.00 \\     
\bottomrule
\toprule
{\bf Scenario B} & & Green & Yellow & {\bf Red} \\
\midrule
Traditional & $\VaR_{0.01}$              & 89.33 & 10.67 & \hphantom{0}0.00 \\
Traditional & $\ES_{0.025}$              & 93.80 & \hphantom{0}6.18 & \hphantom{0}0.02 \\
Comparative & $\VaR_{0.01}$              & \hphantom{0}0.00 & 11.77 & 88.23 \\
Comparative & $(\VaR_{0.025},\ES_{0.025})$ & \hphantom{0}0.00 & 12.78 & 87.22 \\
\bottomrule
\end{tabular}
\end{center}
\end{table}

Table \ref{tab:1} summarizes the simulation results under Scenario A
and B, respectively.  The traditional backtests are performed for the
internal model in the scenario at hand.  Under Scenario A, the four
tests give broadly equivalent results.  The benefits of the
comparative approach become apparent under Scenario B, where the
traditional approach yields highly undesirable decisions in accepting
a simplistic internal model, while a more informative standard model
would be available.  This can neither be in banks' nor in regulators'
interests.  We emphasize that this problem will arise with any
traditional backtest, as a traditional backtest assesses optimality
only with respect to the information used for providing the risk
measure estimates.

Comparative tests based on test statistics of the form $T_2$ in
\eqref{eq:DM} can be used to compare forecasts in the form of full
predictive distributions, provided a proper scoring rule is used
\citep{GneitingRaftery2007}, or to compare risk assessments, provided
the risk measure admits a strictly consistent scoring function, so
elicitability is crucial.  In particular, proper scoring rules and
consistent scoring functions are sensitive to increasing information
utilized for prediction; see \citet{HolzmannEulert2014}.  However, as
consistent scoring functions are not unique, a question of prime
practical interest is which functions ought to be used in regulatory
settings or internally.

Arguably, now may be the time to revisit and investigate fundamental
statistical issues in banking supervision.  Chances are that
comparative backtests, where a bank's internal risk model is held
accountable relative to an agreed-upon standardised approach, turn out
to be beneficial to all stakeholders, including banks, regulators, and
society at large.

\section*{Acknowledgements}

We thank Paul Embrechts, Fernando Fasciati, Fabian Kr\"uger, Alexander
McNeil, Alexander Schied, Patrick Schmidt, and the organisor, Imre
Kondor, and participants of the ``International Workshop on Systemic
Risk and Regulatory Market Risk Measures'' in Pullach for inspiring
discussions and helpful comments.  Tobias Fissler acknowledges funding
by the Swiss National Science Foundation (SNF) via grant 152609, and
Tilmann Gneiting by the European Union Seventh Framework Programme
under grant agreement no.~290976.

\end{document}